\documentstyle[preprint, aps]{revtex}
\tightenlines
\begin{document}
\flushbottom
\draft
\title{Application of Chaos Induced Near-Resonance Dynamics to Locate
the Global Optimum of Functions}
\author{ Rahul Konnur}
\address{ Tata Research Development and Design Centre, \\
          54B, Hadapsar Industrial Estate, Pune. 411 013. India. }
\date{\today}
\maketitle
\begin{abstract}
The problem of locating the global optimum of functions is studied in
a dynamic setting. The dynamics of simple multistable systems under
the influence of chaotic forcing is investigated. When the magnitude
of the forcing signal decays
slowly, it is shown that the system attains an equilibrium state, which
corresponds to the global optimum of the corresponding multimodal
potential function. The role of bifurcations in facilitating the
approach to the global optimum is discussed.
\end{abstract}
\newpage
The problem of locating the global optimum (specifically the
minimum) ${\bf x=x^*}$, of a continuous, convex function
$f({\bf x})$ of $n \ge 1$ variables ${\bf x}$ is considered.
Solving the equation(s) representing the condition that
the derivatives vanish at the extrema of the function, i.e.
\begin{equation}
\frac{\partial f}{\partial x_i}=0 \ \ \ \ {\rm for} \ \  i=1,2,\cdots,n
\end{equation}
is one among several traditional methods available for solving this
problem. However.
the solutions obtained using almost all of the traditional methods
depend on the choice of the initial guess and often correspond
to a local optimum.

Alternately, the given objective function can be recast as a
system of ordinary differential equations (ODE's).
This is possible if we consider the set of equations (1) 
to be a set representing the equilibrium of a set of ODE's in the
variables $\bf {x}$, i.e.
\begin{equation}
\frac{dx_i}{dt}=-\frac{\partial f}{\partial x_i}
\end{equation}
\noindent with $x_i(t=0)=x_i(0), i=1,2,\cdots,n$.
Depending on the choice of the initial conditions, the equilibrium point
attained by the system will correspond to one of the minima of (1) and
one of these will be the global minimum of the objective function (1).
In what follows, the equilibrium point corresponding to the global
minimum will be referred to as the desired equilibrium point.
Even for relatively simple problems, the
initial conditions which guarantee an approach to the desired equilibrium point
are often difficult to estimate.

Here we study the dynamics of (2) when it is subjected to 
chaotic forcing. Specifically, we focus on the case where the
magnitude of forcing is reduced in a gradual manner. Our objective is to
investigate the effect of the magnitude of forcing and the rate of its decay on
the transient dynamics and the eventual equilibrium point attained by (2).
The system we study is
\begin{equation}
\frac{dx_i}{dt}=-\frac{\partial f}{\partial x_i} + C_ie^{-k_2t}
\ \ \ \ {\rm for}\ \ i=1,2,\cdots,n
\end{equation}
\noindent where $C_i$ is the chaotic forcing term and $k_2$ controls the rate
of decay of the forcing. The Lorenz equation with parameters
$\sigma=10; r=60; b=8/3$ has been used to generate the chaotic signal.

In this article we investigate the evolution of simple multistable systems
under the influence of chaotic forcing.
The main result is that with appropriate tuning of the chaotic forcing signal,
the system attains an equilibrium state, which
corresponds to the global optimum of the corresponding multimodal potential
function.
This formulation can be considered to
be a deterministic analog of the simulated annealing optimization
technique \cite{ref10},
with $C_i$ and $k_2$ being analogous to the temperature and the rate of
cooling, respectively.

The optimization problem (1) is thus transformed to a form
which has been studied in the context of stochastic resonance (SR)
\cite{ref1,ref2,ref3,ref4,ref5,ref6,ref7,ref8,ref9,ref9a}.
SR refers to the phenomenon wherein the response of a randomly
perturbed nonlinear system to a very weak periodic signal is significantly
increased with appropriate tuning of the noise intensity.
The mechanism of SR was initially introduced as a possible explanation
of the long time climate changes. In the past few years, this
phenomenon has been discovered and applied to several situations in physics (in
a ring laser and a variety of other bistable systems) and biology (neurological
processes).  A typical example is
in the evolution of the following system
\begin{equation}
\frac{dx}{dt}=\frac{\partial{V(x,t)}}{\partial x} + \sigma \eta
\end{equation}
\noindent where $\eta$ is white noise, $\sigma$ is its magnitude and $V$ is a
time-periodic double well potential
\begin{equation}
V(x,t)= -\frac{x^4}{4}+\frac{x^2}{2}+Ax\cos(\omega t)
\end{equation}

Two time scales characterize the evolution of this system. The first corresponds
to oscillations in one of the two potential wells (intra-well dynamics), while
the second corresponds to oscillations between the wells (inter-well dynamics).
The latter characterizes the mean time of barrier crossings and accounts for the
global dynamics of noise induced transitions between the potential wells. It has
to be noted that the amplitude of the periodic modulation signal is sufficiently
small in the sense that the system cannot switch between the potential wells in
the absence of noise. For the example system given above, it has been shown
that there exist a range of values of $A$, $\sigma$ and $\omega$ where the
jumps between the two oscillating wells are strongly synchronized, as a
consequence of a {\it resonance} between the periodic forcing and random
perturbation.

In analogy with SR, chaotic resonance (CR) refers to the transitions induced
by chaotic forcing of nonlinear systems \cite{ref12,ref13,ref14,ref15,ref16}.
The presence of deterministic chaos plays the role of random perturbation.
Depending on the magnitude of the chaotic forcing signal, a multistable system
can show intra-well dynamics or inter-well dynamics. This is similar to that
discussed earlier in the context of SR. Thus under appropriate conditions, the
dynamics of the system can be strongly synchronized with the time variation of
the periodically varying control parameter. In several systems, CR has been
shown to occur near the onset of chaos and near crisis bifurcation points.
Briefly, a crisis bifurcation point refers to the bifurcations following which
a system undergoes a transition from intra-well to inter-well dynamics
\cite{ref17}.

Inducing a crisis bifurcation is
therefore the key step in our objective of locating the global optimum of
functions.
The proposed scheme involves inducing crisis bifurcations in the system
by driving the system with an appropriate amount of a chaotic input.
We have verified that signals
from a variety of chaotic systems can be used to induce crisis bifurcations
in several multistable systems of interest to us. Transitions from
intra-well to inter-well dynamics can be easily obtained by increasing the
magnitude of the forcing. Once the system starts to show inter-well dynamic
behavior, the magnitude of the forcing signal is reduced in a gradual manner.
An important result of this work is that for a vast choice of initial
conditions. the system (2) stabilizes at the state
which corresponds to the global minimum of the potential function $f({\bf x})$.
The first of the four example minimization problems  presented below
has been selected with the purpose of illustrating the nature of the
bifurcations which
make it possible for the system to reach the global minimum. The remaining three
examples are well known test functions and have been selected to show the
validity of the technique.

The first example is:
\begin{equation}
{\rm min} \ \  f(x) = \frac{x^4}{4} - 16x^2 + 5x
\end{equation}
The ODE system which is solved for determining the global minimum is:
\begin{equation}
\dot x = -f^{\prime}(x) + k_1 c e^{-k_2 t};  \ \ \ f^{\prime}(x)=
(x^3 - 32x + 5)
\end{equation}
Here $c$, $k_1$ and $k_2$ represent the chaotic forcing signal,
its magnitude and the decay constanT,
respectively. The negative sign is needed to ensure that
the two minima of (6) are the stable equilibrium points of (7).

When $k_1=0$, the unforced system (7) has an unstable equilibrium point
near $x=0.40$ and this corresponds to the maximum value of (6).
This equilibrium acts as a {\it potential barrier} and prevents
an approach to the negative equilibrium point for almost all choices of 
initial conditions that are positive.
The crossover from the local to the global
minimum can occur only if the forcing input is sufficient to overcome the
potential barrier associated with the presence of this maximum. For small
magnitude of forcing and depending on the initial condition, the system (7)
oscillates in the vicinity of one of the two stable equilibrium points.
As the magnitude is increased past a critical value, the oscillatory state
associated with the positive equilibrium point disappears.
This is due to the occurrence of a 
crisis bifurcation
This value of the magnitude of forcing is the least value which makes it
possible to reach the desired equilibrium state.
Following this bifurcation and irrespective of the initial conditions,
the system (7) intermittently evolves across its
former basin of attraction, which no longer exists. Now the only stable
state is that associated with the negative equilibrium point.
As the magnitude of forcing is gradually reduced, the system spends more time
in the vicinity of this equilibrium point and hence has an increased probability
of eventually stabilizing there.

Fig 1(a) and 1(b) show the dynamics of the system before it reaches the
local equilibrium (Fig. 1(a)) and global equilibrium (Fig. 1(b)) points. As
can be seen, the system oscillates near each minima before stabilizing at the
global minimum. For sufficiently low decay rates,
the systems always reaches the negative equilibrium solution, which is the
global minimum of (6). Finally, it must be noted that the magnitudes of forcing
and the decay constant are the crucial parameters which together decide whether
or not the system actually reaches the desired state.

To summarize, the ability of this system (a particle in a double-well
potential) to
jump over the {\it energy barrier} is acquired following the occurrence of a
{\it  crisis bifurcation}. 
Increased sensitivity of the system following the occurrence of a crisis
bifurcation (near-resonance dynamics)
allows stabilization the system at a state
corresponding to the global minimum of the function.

The second example is a slightly more complex function:
\begin{equation}
{\rm min} \  \ f(x) = (x_1^2+x_2-11)^2+(x_1+x_2^2-7)^2+
(x-3)^2+(x-2)^2
\end{equation}
This function has four minima with the
global minima at $(x_1=3, x_2=2)$. We solve the following equation
\begin{eqnarray}
\frac{dx_1}{dt}&=&-(4x_1^3 + 2x_2^2 + 4x_1x_2 - 40x_1 - 20) + k_1 c e^{-k_2t} \\
\frac{dx_2}{dt}&=&-(4x_2^3 + 2x_1^2 + 4x_1x_2 - 24x_2 - 26) + k_1 c e^{-k_2t}
\end{eqnarray}

The terms within the braces in (9) and (10) are the derivatives with respect to
the dependent variables $x_1$ and $x_2$ respectively.
For low decay rates $(k_2<0.80)$ and when $(k_1>1.50)$, the solutions of
these equations evolve (randomly) between the different minima before finally 
stabilizing at the state corresponding to the global minimum (Fig. 2).
This again shows the ease with which the global minimum of 
multistable systems can be located.

Fig. 3 shows results of a parametric study carried out in $k_1$-$k_2$ space.
1000 sets of randomly chosen initial conditions were used and the number of
initial conditions leading to the global minimum was computed. This figure
shows that for sufficiently high values of $k_1$ and sufficiently low values of
$k_2$, the system always stabilizes at the desired equilibrium point.

The third example considered is that of the multimodal Rastrigin function with
one and three variables
\begin{equation}
{\rm min}\ \ f(x)=10n + \sum_{i=1}^{n}(x_i^2-10 \cos(2\pi x_i)),\ n=1
\ {\rm or}\ 3
\end{equation}
\noindent with a global minimum at $x_i=0,i=1,\cdots, n$.
Fig. 4 shows the evolution of the forced system and the approach to the desired
equilibrium point. Values of $k_1>20$ and $k_2<0.005$ were found to be
sufficient to stabilize the system at the desired equilibrium point.

A common feature of the functions considered previously is that the global
minima lies in the deepest potential well. In such systems, addition
of a chaotic forcing signal
systematically destroys the local minimum in the order of increasing depth of
potential wells. This makes it possible to locate the global minimum of the
functions. A similar argument shows that it will not be possible to reach the
global minimum using the scheme (3) for systems in which the global minimum
and the deepest potential well
do not coincide. An example of such a situation is the function (Fig. 5)
\begin{equation}
{\rm min} \ \ f(x) = 10x(x-0.2)(x-1)(x-2)(x-3)(x-4)+3x^3+x^2
\end{equation}

In such cases, on/off tuning of the chaotic forcing signal using a suitably
chosen acceptance or rejection criterion based on the magnitude of the function
(12) provides a way of locating the global minimum. Fig. 6 shows a typical
example of the temporal evolution of the dynamical system obtained from (12).

In conclusion, it has been shown that it is possible to locate the global
optimum of convex functions by solving the problem in a dynamic setting.
Chaotic forcing of appropriate magnitudes lead to the occurrence of crisis
bifurcation(s) in the multimodal system. This enables the system to sequentially
overcome the (local) energy barriers.  A suitably low rate of decrease of
the forcing signal stabilizes the system at the state in which its
energy is the lowest.

\newpage
\begin{center}
\bf \large {LIST OF FIGURES}
\end{center}
FIG. 1. Evolution of the system (7) with time showing the approach
to the (a) local minimum for $k_1=2.750$, $k_2=0.01$; (b) global minimum for
$k_1=5$, $k_2=0.01$. The chaotic drive is the $y$ variable of the Lorenz
equation. The dashed lines show the location of each of the two stable
equilibrium points.

\noindent FIG. 2. Evolution of $x_1$ and $x_2$ of Equation (9) and (10)
with time showing the approach
to the global minimum, $k_1=5$ and $k_2=0.05$. The chaotic drive is the $y$
variable of the Lorenz equation.
The dashed lines show the location of the desired equilibrium point.

\noindent FIG. 3. Parametric dependence of the number fraction of initial
conditions $F$ stabilizing at the local minima of the Himmelblau function (8).

\noindent FIG. 4. Temporal evolution of the dynamic system obtained from
(11) showing the approach to the global minimum for
$k_1=25$, $k_2=0.01$. The chaotic drive is the $y$ variable of the Lorenz
equation.  The dashed line shows the location of the desired equilibrium point.

\noindent FIG. 5. Graphical representation of the function (12). This is
an example of the case where the
global minimum does not coincide with the deepest potential well.

\noindent FIG. 6. Temporal evolution of the dynamic system obtained from
(12) showing the approach to the global minimum for
$k_1=10$, $k_2=0.008$. The simulation was carried out using an on/off control
strategy. The chaotic drive is the $y$ variable of the Lorenz equation.
The dashed line shows the location of the three stable equilibrium points.
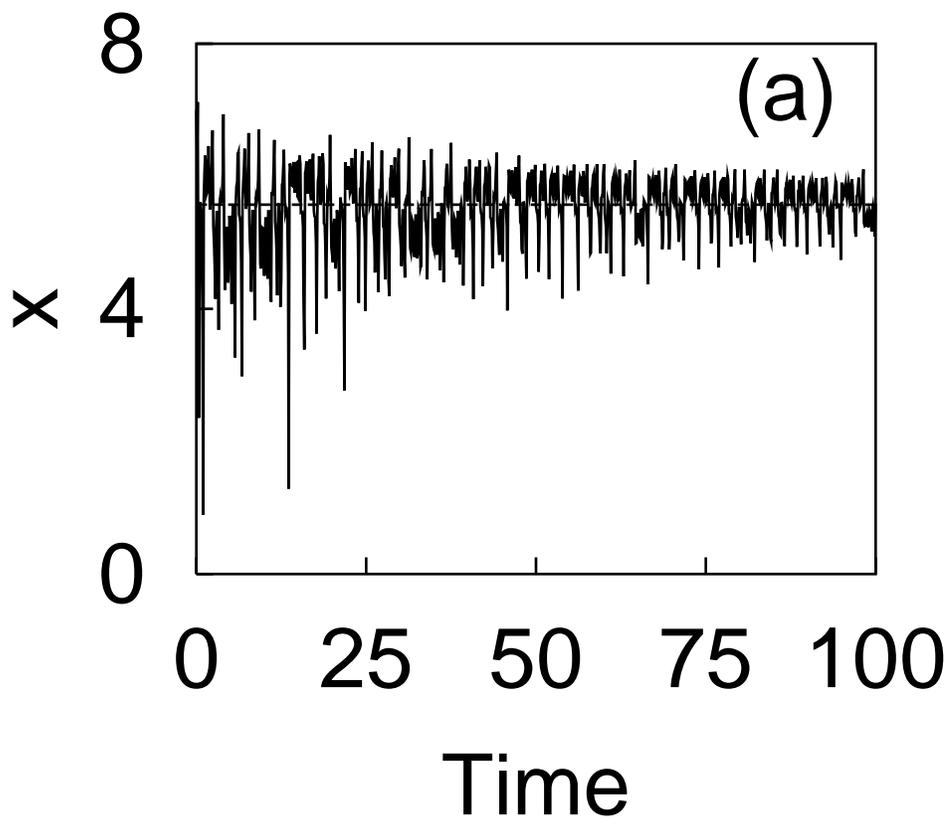
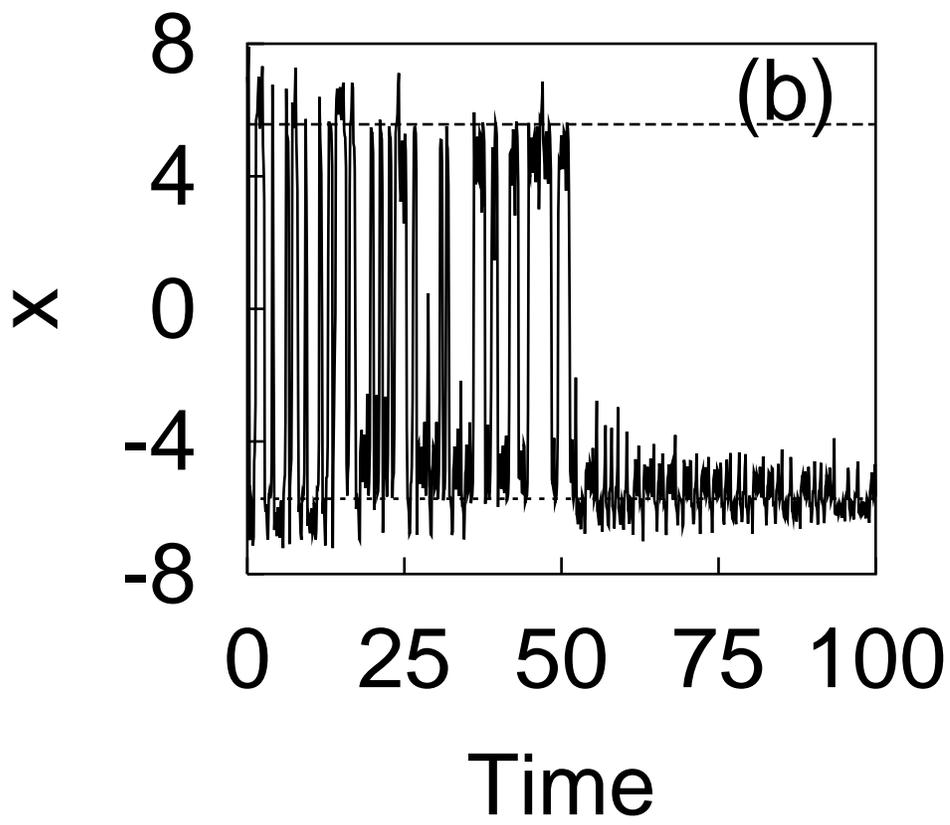
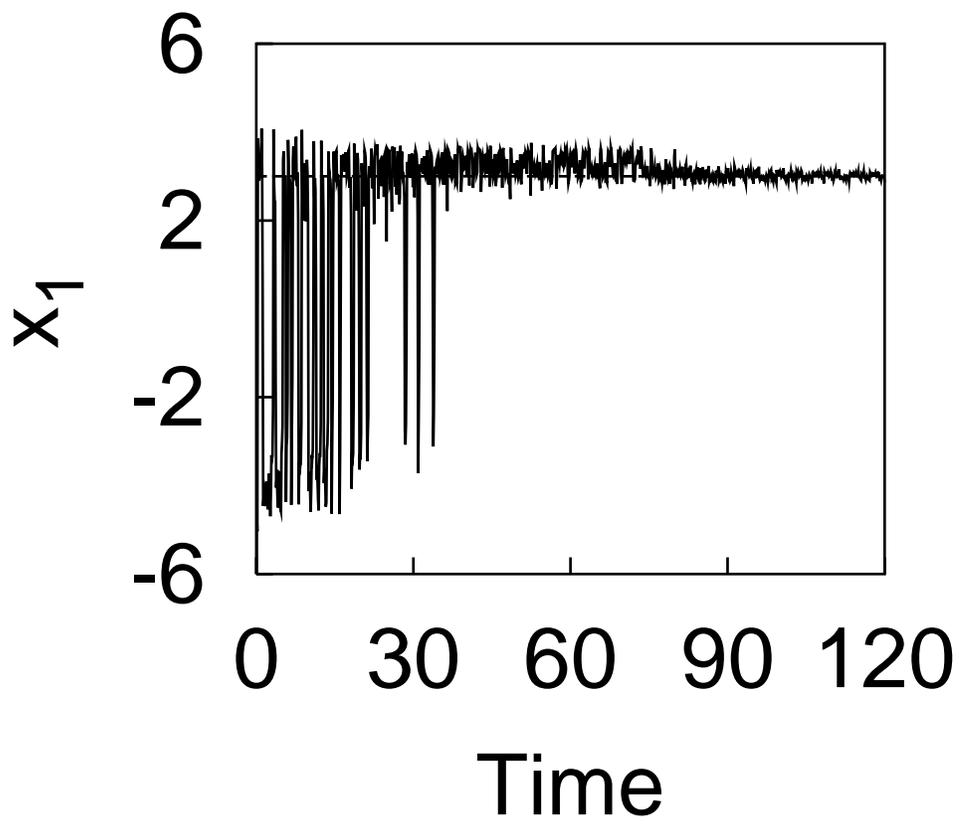
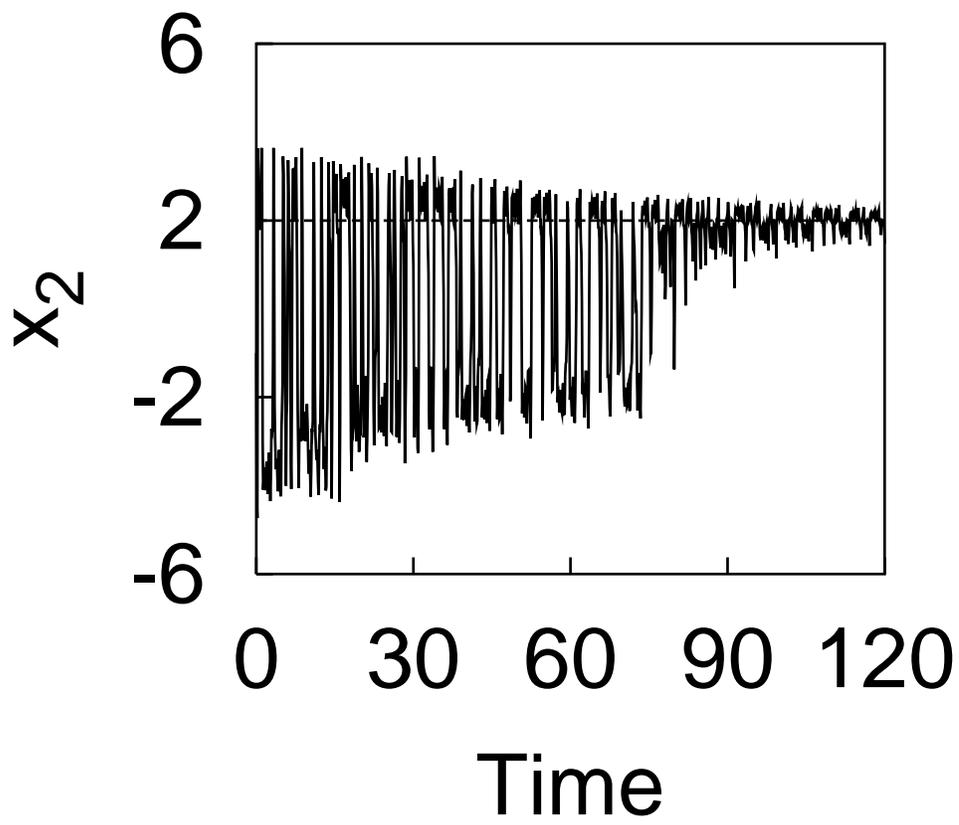
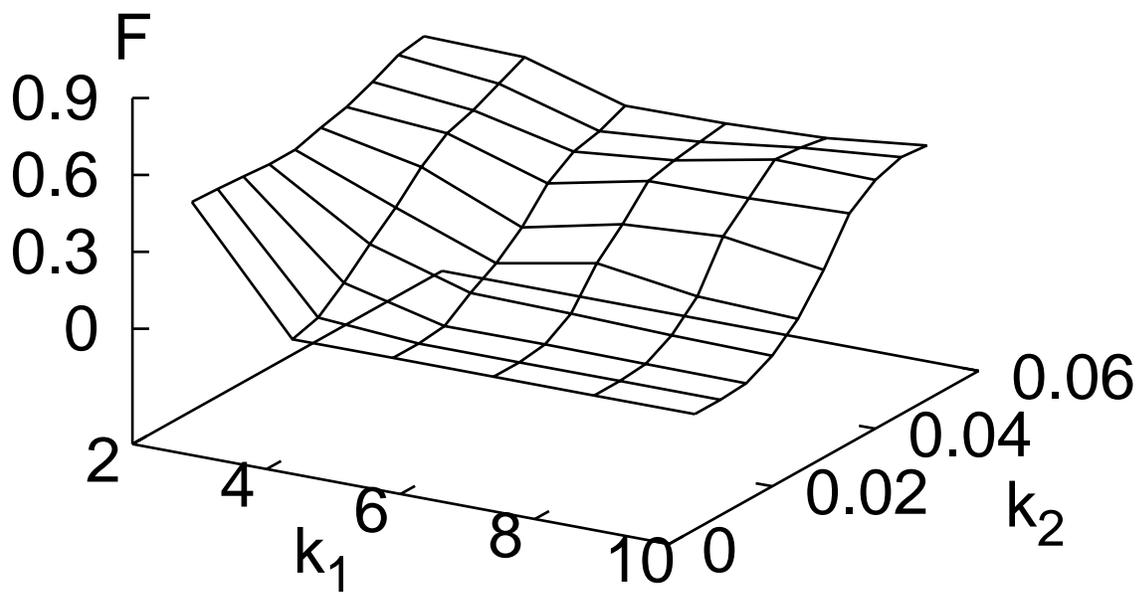
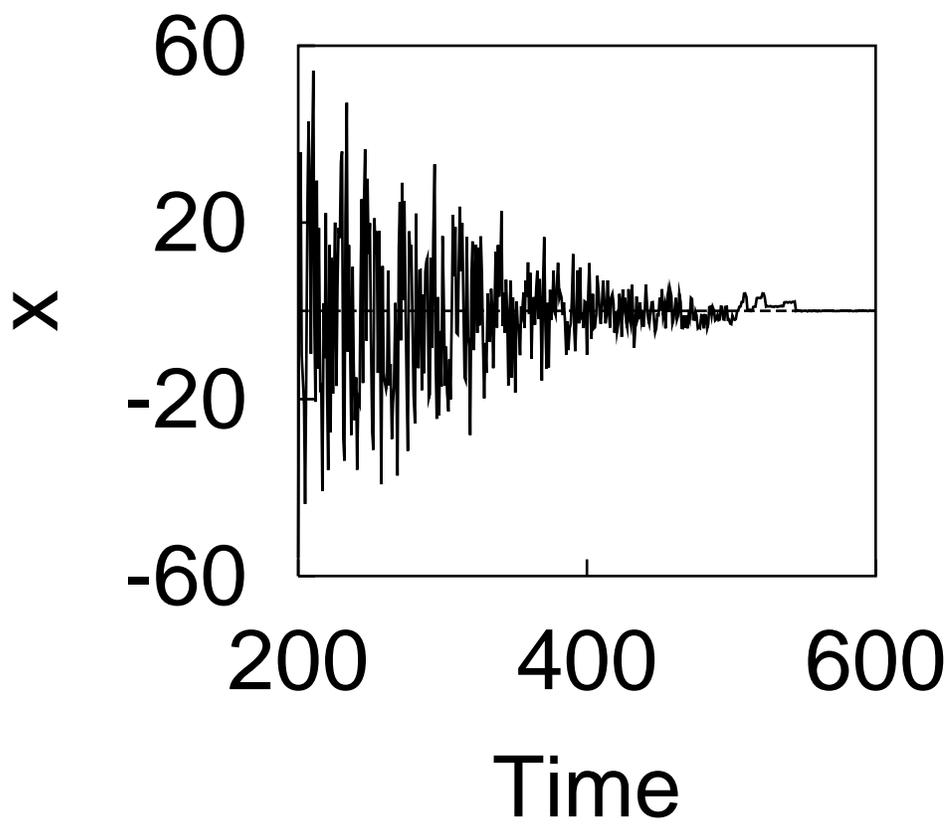
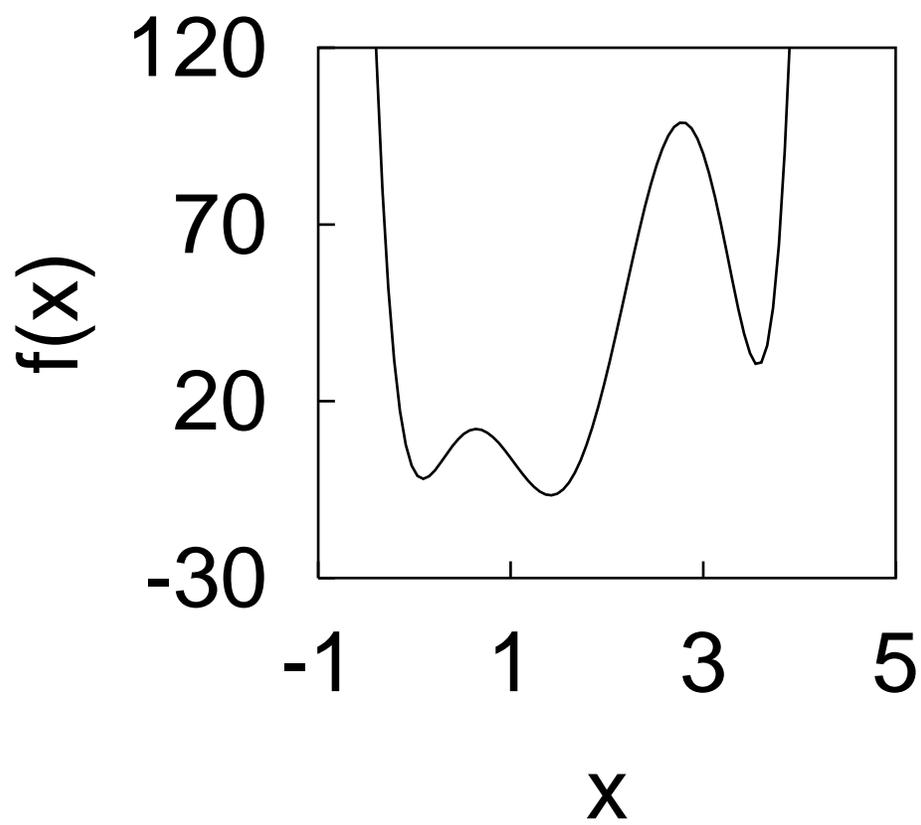
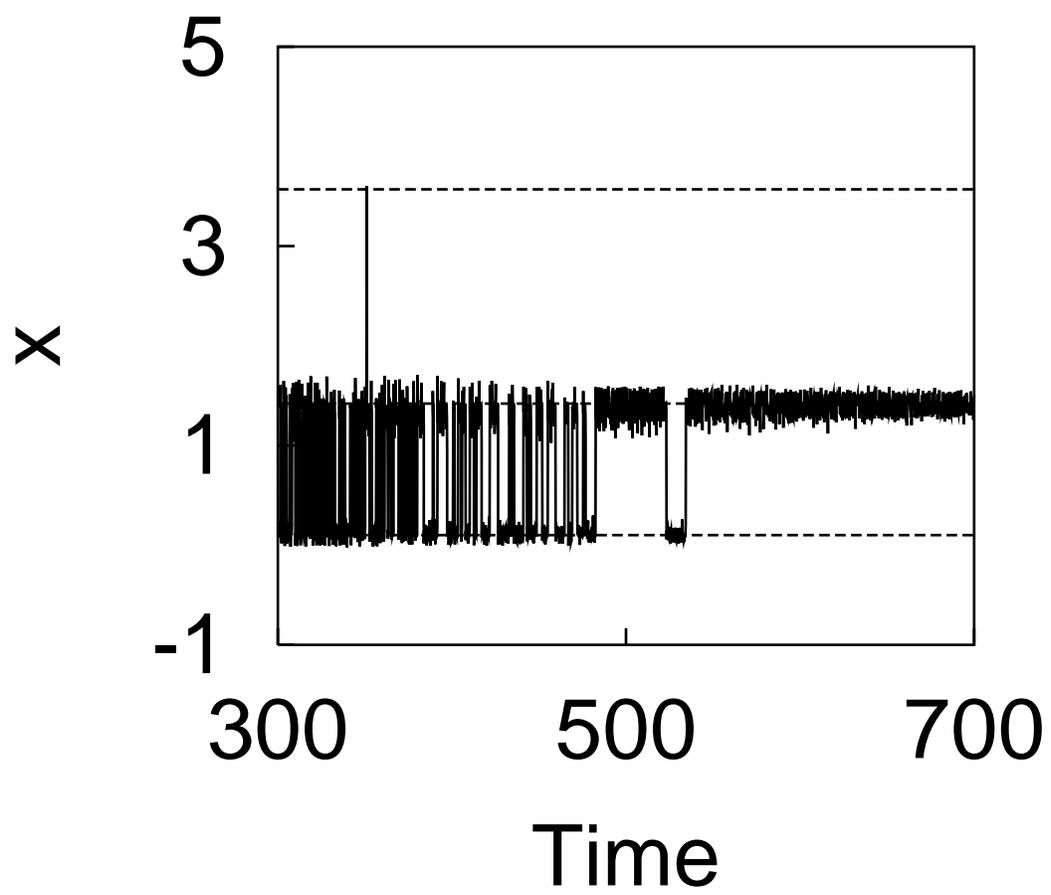
\end{document}